\documentclass[letterpaper,twocolumn,10pt]{article}
\usepackage{usenix2019,epsfig,endnotes}

\usepackage[utf8]{inputenc}
\usepackage{colortbl}
\usepackage{xcolor}
\usepackage{graphicx}
\usepackage{subcaption}
\usepackage{verbatim}
\usepackage{url}
\usepackage[draft=true]{minted}
\usepackage{lipsum}
\usepackage{amsthm}
\usepackage{mathtools}
\usepackage{xfrac}
\usepackage[export]{adjustbox}
\usepackage{longtable}
\usepackage{booktabs}
\usepackage[section]{placeins}
\usepackage{hyperref}
\usepackage{comment}
\usepackage{pgf-umlsd}

\newcommand\todo[1]{\textcolor{red}{#1}}

\begin{document}
\date{}

\title{Volatile memory forensics for the Robot Operating System} 


\author{
{\rm Víctor Mayoral Vilches},
{\rm Laura Alzola Kirschgens}, {\rm Endika Gil-Uriarte}, {\rm Alejandro Hernández}\\
ALIAS ROBOTICS,\\
Calle Venta de la Estrella 6, Pab 110\\ 
01006 Vitoria-Gasteiz, Álava, Spain
\and
{\rm Bernhard Dieber}\\
JOANNEUM RESEARCH\\
Institute for Robotics and Mechatronics\\
Lakeside B08\\
9020 Klagenfurt, Austria
} 

\maketitle

\thispagestyle{empty}

\subsection*{Abstract}
The increasing impact of robotics on industry and on society will unavoidably lead to the involvement of robots in incidents and mishaps. In such cases, forensic analyses are key techniques to provide useful evidence on what happened, and try to prevent future incidents. This article discusses volatile memory forensics for the Robot Operating System (ROS). The authors start by providing a general overview of forensic techniques in robotics and then present a robotics-specific Volatility plugin named \emph{linux\_rosnode}, packaged within the \emph{ros\_volatility} project and aimed to extract evidence from robot's volatile memory. They demonstrate how this plugin can be used to detect a specific attack pattern on ROS, where a publisher node is unregistered externally, leading to denial of service and disruption of robotic behaviors. Step-by-step, common practices are introduced for performing forensic analysis and several techniques to capture memory are described. The authors finalize by introducing some future remarks while providing references to reproduce their work.


\section{Introduction}

As robots become widespread and used in a variety of different applications, it is expected that malicious actors will start  compromising robotic systems for their own purposes. Potential misuse of robots may range from privacy invasion by unauthorized remote access to sensors (like the camera of your robot at home) up to causing damage or harm to the robot's environment or persons in its vicinity. In an attempt to diagnose and prevent robot-aided crime, robot forensics proposes a number of scientific tests and methods to obtain, preserve and document evidence from robot-related crimes. \\
\newline
Robot forensics are closely connected to computer forensics as a subgroup of digital forensics, namely summed up as Digital Forensic Incident Response competences (DFIR). This entails the interaction of crime, evidence, science and law. According to Abeykoon and Feng\cite{abeykoon2017forensic}, the field of computer forensics is young and was initiated around 1970 in the United States, in response to the requirement for more extensive access to solve cyber-criminal activities related to protocols in a computer. Within the digital forensics community, there are different areas of interest. Of most relevance are: a) non-volatile memory forensics, which includes hard drives and storage peripherals, b) network forensics and c) volatile (RAM) memory forensics. For the purpose of robotics, the authors classify these areas in two subgroups:
\begin{itemize}
    \item Memory forensics
    \item Networking forensics
\end{itemize}

\noindent In this article, the authors focus on robot memory forensics that could become the stepping stone towards the creation of future Robot-specific Forensics and Incident Response (RFIR) teams. In particular, volatile (RAM) memory forensics is studied for its interest in robotics. The goal is to perform an online analysis of a robot's memory to detect potential manipulations in the running system.\\
\newline
RAM forensics aims to use memory management structures in computers such as arrays, bitmaps, records, linked lists, hash tables, trees, etc, to extract files or other information resident in a computer’s physical memory. These files can later be used to prove that a crime has transpired, or to trace how it came to pass. In robotics, it is relevant to note that RAM typically presents all the recent information that runs across all devices of the robotic system. This becomes specially interesting given the popularity of distributed architectures in robotics. Regardless of the availability of non-volatile memory, most robot components, as participants of the robot network, may contain relevant information in their RAM memory. Specially, most robots have a dedicated component for computation or cognition. Such component is in charge of concentrating recent samples from sensors and actuators to reason and react accordingly. During this study, the authors will focus on such a component, namely on the cognition device of the robot. Moreover, the focus will be on the study of the Robot Operating System (ROS)\cite{quigley2009ros}, which to many is the most popular framework for robot application development.\\
\newline
Besides ROS, the authors also restrict themselves to the study of forensics in Linux-based robotic systems. The rationale behind this assumption is twofold. First, ROS was traditionally only supported in Linux, which made the community grow in that direction. Nowadays, most ROS robotic systems are powered by Linux or by a POSIX-compliant Unix variant. Second, a growing number of robot components run embedded Linux, in which memory-only file systems are used. Such artifacts are lost when the device is powered down. Thus, as pointed out by Ligh et al. \cite{ligh2014art}, in many cases, preserving RAM is the best (and sometimes the only) method to determine which files an attacker accesses.\\
\newline
The sections below are structured as follows: section \ref{prior} provides an overview of prior art in the area of robot forensics and particularly, robot memory forensics. Section \ref{forensics} discusses how to perform a memory forensics analysis in ROS, including memory acquisition, tools, and presenting a walk-through on several use cases. Finally, section \ref{conclusions} presents the conclusions and future work. In addition, Appendix \ref{forensicsvsreversing} clarifies the differences between forensics and reversing.\\

\section{Previous work}
\label{prior}


There are very few studies that aim to perform forensic research in robots. However, as they are digital systems, much of the existing literature applies to them too. Within previous work, a large share of attention has been paid to analyze non-volatile media such as hard drives or storage peripherals, also present in robots. More recently, the rise of networks has created an interest in the study of network-based evidence. Both of these subjects have existing, extensive bodies of knowledge, as exemplified by Luttgens et al.\cite{luttgens2014incident} and Cichonski et al.\cite{nistcomputersecurity}.\\
\newline
This is not the case of RAM memory. Few studies focus on volatile memory forensics and even less are relevant or applicable to robotics. Some of the most interesting include the talk given by Mariusz Burdach at Black Hat 2016\cite{burdach2006physical}, the popular Android's memory acquisition and analysis\cite{sylve2012acquisition} or \cite{ligh2014art}, where a comprehensive discussion on forensic techniques for volatile memory is held for different operating systems.\\
\newline
A recent study on memory forensics for ROS was conducted by Abeykoon et al.\cite{abeykoon2017forensic}, where the authors claim they \emph{present the first methodology and toolset for acquisition and deep analysis of volatile physical memory from robot operating system devices}. This article presents certain information on a somewhat interesting manner, but lacks on formal methods, and the content presented is not useful for the expert reader. Within the text, the Robot Operating System is presented as a relatively new concept, although ROS was launched first in 2007. At the time of writing, the commonly cited ROS article, published in 2009, has received about 4959 citations. The ROS community has spread across countries and events dedicated to ROS are appearing in different geographical areas. Regardless of certain discrepancies with the understanding of ROS, Abeykoon et al. provide a nice introduction to the field of forensics and make use of popular forensic tools such as Volatility\cite{walters2007volatility} or LiME\cite{sylve2012lime}. The authors, however, do not present any novel contribution that justifies their claims, nor any description to reproduce their work. \\
\newline
In an attempt to provide a foundational path for the growth of robot forensics, and specifically aimed at volatile memory, the content below presents a study on ROS volatile memory forensics. The authors introduce common practices for performing forensic analysis in robots powered by ROS and describe several techniques to capture memory. They discuss the tools employed, introduce their contributions, and explain how to use them through a study case that provides a walk-through on the forensics of a ROS system. To finalize, the authors introduce some remarks and insights for future research and provide references to reproduce their work.

\section{ROS volatile memory forensics}
\label{forensics}

\subsection{Volatile memory acquisition in robots}

As pointed out before, several techniques from traditional digital forensics will be useful for robots. Among them, memory acquisition for robots can be performed in a similar manner. The following subsections will describe some of the most popular techniques applicable to robots:

\subsubsection{Using \texttt{/dev/mem}}

\texttt{/dev/mem} provides access to the physical memory of a running system. According to Ligh et al.\cite{ligh2014art}, \texttt{/dev/mem} was the  most popular interface for memory acquisition before being disabled in most Linux distributions for security reasons. Currently, its access remains restricted\cite{devmemrestricted} in popular distributions like Ubuntu through the \texttt{CONFIG\_STRICT\_DEVMEM} kernel option.\\
\newline
Still, with the right privileges, a forensics analyst should be able to get the right level of permission. Unfortunately, this interface only allows to access the first 896MB of RAM memory with commands like the ones in Listing \ref{listing:captmem1}. The limitation of this interface gave birth to other methods.

\begin{listing}[ht]
\caption{{Command line to capture volatile memory with \texttt{/dev/mem}}}
\begin{minted}
[
frame=lines,
framesep=2mm,
baselinestretch=1.2,
fontsize=\footnotesize,
linenos
]
{bash}
sudo dd if=/dev/mem of=/home/vagrant/robot.dd 
    bs=1MB count=10
\end{minted}    
\label{listing:captmem1}
\end{listing}

\subsubsection{Using \texttt{ptrace}}
\texttt{ptrace} is the userland debugging interface that Linux provides and, according to \cite{ligh2014art}, it is not suitable for robust memory acquisition. It can acquire pages from running processes without any changes required in the kernel. This makes the process of acquiring memory less aggressive and simplified in certain robots. However the information acquired keeps solely restricted to process-related pages.\\
\newline
An interesting tool to capture memory in this manner is \url{https://github.com/citypw/lcamtuf-memfetch}. Similar to what is described in \cite{ligh2014art}, this repository offers an applications that, when compiled and executed, reads the starting and ending addresses of process' memory ranges from \texttt{/proc/<pid>/maps} and then uses \texttt{ptrace} to dump each page to disk.\\
\newline
The previous two methods were widely used on 32-bit systems, yet with the growth of 64-bit devices, new acquisition methods appeared that made use of kernel modules to access the complete range of RAM memory. The following two are, as of today, the most popular in the Linux world.

\subsubsection{Using \texttt{/dev/fmem}}

\texttt{/dev/fmem} was created as an extension of \texttt{/dev/mem}. The character device appears by loading a kernel driver and it exports physical memory for other programs to access while providing a number of advantages as described at \cite{ligh2014art}. Listing \ref{listing:captmem2} shows how to install the character device and \ref{listing:captmem3}, how to use it to acquire memory.

\begin{listing}[ht]
\caption{{Installing \texttt{/dev/fmem} character device for memory acquisition}}
\begin{minted}
[
frame=lines,
framesep=2mm,
baselinestretch=1.2,
fontsize=\footnotesize,
linenos
]
{bash}
git clone https://github.com/NateBrune/fmem
cd fmem && make
sudo ./run.sh
\end{minted}    
\label{listing:captmem2}
\end{listing}

\begin{listing}[ht]
\caption{{Command line to capture volatile memory with \texttt{/dev/fmem}}}
\begin{minted}
[
frame=lines,
framesep=2mm,
baselinestretch=1.2,
fontsize=\footnotesize,
linenos
]
{bash}
sudo dd if=/dev/fmem of=memory.dump 
    bs=1MB count=1000
\end{minted}    
\label{listing:captmem3}
\end{listing}

\subsubsection{Using the \textit{Li}nux \textit{M}emory \textit{E}xtractor (LiME)}
LiME is often presented\cite{ligh2014art} as the latest Linux memory acquisition tool. It is operated by loading a kernel driver that, instead of creating a userland accessible character device (like \texttt{/dev/fmem} above), fetches the memory from the kernelspace. This enhances the accuracy of the resulting samples due to the removal of context switches between kernel and user spaces for transferring data.\\
\newline
LiME allows to capture memory in different formats. Listing \ref{listing:captmem4} shows how to install LiME in the \texttt{\$HOME} directory, \ref{listing:captmem5} how to use it and \ref{listing:captmem6} how to remove the kernel module to cleanup after the memory capture. LiME will be the tool used throughout the analysis.–

\begin{listing}[ht]
\caption{{Installing LiME in the \texttt{\$HOME} directory for memory acquisition}}
\begin{minted}
[
frame=lines,
framesep=2mm,
baselinestretch=1.2,
fontsize=\footnotesize,
linenos
]
{bash}
cd $HOME && git clone https://github.com/504ensicsLabs/LiME
cd $HOME/LiME/src && make
cd $HOME/LiME/src && cp lime-*.ko lime.ko
cd $HOME/LiME/src && sudo mv lime.ko /lib/modules/
\end{minted}    
\label{listing:captmem4}
\end{listing}

\begin{listing}[ht]
\caption{{Command line to capture volatile memory with LiME through the kernel module}}
\begin{minted}
[
frame=lines,
framesep=2mm,
baselinestretch=1.2,
fontsize=\footnotesize,
linenos
]
{bash}
sudo insmod /lib/modules/lime.ko 
    "path=/home/vagrant/robot.lime 
    format=lime"
\end{minted}    
\label{listing:captmem5}
\end{listing}

\begin{listing}[ht]
\caption{{Command line to remove LiME's kernel module}}
\begin{minted}
[
frame=lines,
framesep=2mm,
baselinestretch=1.2,
fontsize=\footnotesize,
linenos
]
{bash}
sudo rmmod /lib/modules/lime.ko
\end{minted}    
\label{listing:captmem6}
\end{listing}


\subsection{Volatility and the \emph{ros\_volatility} plugins}

The Volatility framework \cite{walters2007volatility} is an open collection of tools for the extraction of digital artifacts from volatile memory (RAM) samples. It is written in Python, licensed under GPLv2 and provides a wide variety of algorithms that run in Linux, Mac OS or Windows. The framework consists of several subsystems that work together to provide a robust set of features. Among these subsystems is the \emph{plugin system}. Plugins allow you to extend the Volatility framework with new capabilities. For example, an address space plugin could introduce support for operating systems that run on new CPU chipsets. Volatility is indeed a fantastic tool. However, currently and according to its README, its support in Linux is limited, and only kernels up to 3.6, which imposes certain limitations for robots running modern kernels.\\
\newline
The authors will be using Volatility and extend it with a new set of plugins named \emph{ros\_volatility}\endnote{\url{https://github.com/aliasrobotics/ros_volatility}} that have been developed to analyze the memory of ROS-related artifacts. At the time of writing \emph{ros\_volatility} presents only a single plugin:
\begin{itemize}
    \item \textbf{linux\_rosnode}: A basic class used to fetch all ROS nodes from memory. It extends the linux\_pslist class to obtain all processes and filters according to those that make use of the typical ROS libraries. In addition, it checks
    the sockets of each one of the nodes and verifies whether the node was or wasn't registered in the ROS network.
\end{itemize}
Additional plugins are being developed and contributed to analyze memory and determine issues in topics and other abstractions of the ROS ecosystem.

\subsection{A Study case: Unauthenticated registration/unregistration of ROS Nodes}
Described by Dieber et al.\cite{dieber2017security}, the ROS Master API\endnote{\url{http://wiki.ros.org/ROS/Master_API\#register.2BAC8-unregister_methods}} requires no authentication capabilities to register and unregister publishers, subscribers and services. This leads to a reported vulnerability\endnote{\url{https://github.com/aliasrobotics/RVDP/issues/87}} that can easily be exploited with off-the-shelf penetration testing tools, provided an attacker has access to the internal robot network. Based on this, possible exploits range from eavesdropping via denial of service up to injection of false data.\\
\newline
To assess the severity of the vulnerability, the authors of this paper have used the Robot Vulnerability Scoring System (RVSS)\cite{RVSS} under the following assumptions:
\begin{itemize}
    \item Vectors of attack come from within the internal network of the robot. As discussed by Mayoral et al.\cite{RSF}, most robots separate their networking setup in two different networks: an internal one, where the software and hardware components operate and an external one, meant to access the robot from the outside world. This assumption is not entirely true, given recent discoveries\cite{demarinis2018scanning}. Still, provided some basic security awareness and taking a conservative approach, the authors assume that this vulnerability should apply only to those with access to the robot \emph{internal} network, where the ROS Master is operating.
    \item Attack complexity is low given the availability of off-the-shelf penetration testing tools that can exploit this flaw.
    \item No safety implications have been considered since the vulnerability affects ROS, what can be classified as a (software) component and not a complete robot system by itself. It should be noted, however, that a robotic system using a vulnerable ROS setup could easily cause human harm and thereby imply environmental or even human safety hazards.
\end{itemize}

\noindent Following this conservative approach, the vector of the vulnerability is 
\texttt{\small $RVSS:1.0/AV:IN/AC:L/PR:N/UI:N/Y:T/S:U/C:H/I:N/A:H/H:N$}. Altogether, and as stated by Mayoral et al., this vulnerability scores with \textbf{7.6} out of 10, which implies a \emph{High} degree of severity according to the classification proposed.\\


\noindent The following section will show how an attack can exploit this vulnerability and cause selected nodes to be unregistered from the robot network.



\begin{figure}[htbp]
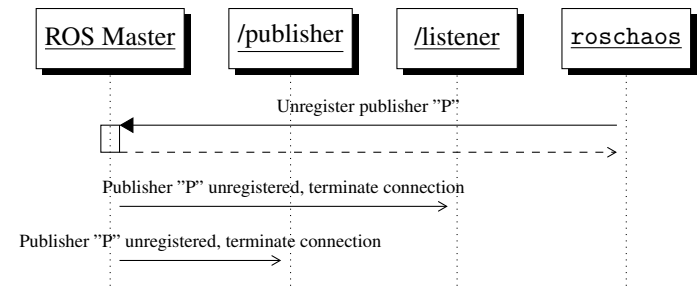

 \centering
 \tikzset{
  every picture/.append style={
    scale=0.2
  }
}
\begin{sequencediagram}
	\renewcommand\unitfactor{1.8}
	\newinst{Master}{ROS Master}
	\newinst[2.9]{Publisher}{/publisher}
	\newinst[2.9]{Subscriber}{/listener}
	\newinst[2.9]{Attacker}{\texttt{roschaos}}
    \begin{scope}[font = \scriptsize]{ }
    	\stepcounter{seqlevel}
		\begin{call}
		    {Attacker}{Unregister publisher "P"}{Master}{}
		\end{call}
		\stepcounter{seqlevel}
		\mess{Master}{Publisher "P" unregistered, terminate connection}{Subscriber}
		\stepcounter{seqlevel}
		\mess{Master}{Publisher "P" unregistered, terminate connection}{Publisher}
		
   \end{scope}
\end{sequencediagram}
 \caption{Sequence diagram of an unauthenticated unregistration attack in ROS.}
 \label{fig:attacks/stealth_publisher}
\end{figure}

\subsubsection{Attacking ROS}
To simulate the attack in an artificial robotic system, the authors made use of the Robotics CTF\cite{rctf} public resources\endnote{\url{https://github.com/aliasrobotics/RCTF}} and constructed an scenario with a robotic system as described in Figure \ref{figure:robot1}. Besides the simulated robot, the scenario also contains different robot penetration testing tools. Mainly \emph{ROSPenTo}\endnote{\url{https://github.com/jr-robotics/ROSPenTo}} and \texttt{roschaos}\endnote{\url{https://github.com/ruffsl/roschaos}}. ROSPenTo is a .NET-based tool that allows for analyzing and manipulating a ROS-network on a very fine-grained level with a high degree of stealthiness. Roschaos is meant for high-impact attacks aiming at causing more damage in a single step.\\
\newline
The application contains two nodes, a \emph{/publisher} node running in a process called 'talker' and a \emph{/listener} node running in a process with that name. Along with that, ROS starts also the roscore process, which contains the ROS master. There is also a ROS node named \emph{/rosout} that is part of the ROS core.
The attack is described in listing \ref{listing:attack1} and the resulting status in listing \ref{listing:attack2}:

\begin{figure}
  \centering
   \includegraphics[width=0.5\textwidth]{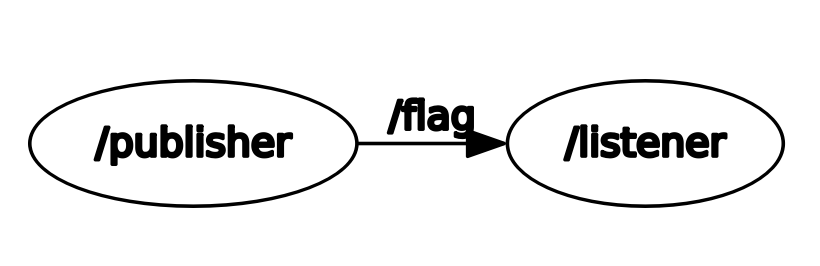}
  \caption{Robot subject of the study case 1: \emph{Unauthenticated registration/unregistration of ROS Nodes}}
  \label{figure:robot1}
\end{figure}

\begin{listing}[ht]
\caption{{Unauthenticated unregistration attack on ROS Master API}}
\begin{minted}
[
frame=lines,
framesep=2mm,
baselinestretch=1.2,
fontsize=\footnotesize,
linenos
]
{bash}
root@655447dc534c:/# rosnode list
/listener
/publisher
/rosout

root@655447dc534c:/# roschaos master unregister 
    node --node_name /publisher
Unregistering /publisher
\end{minted}    
\label{listing:attack1}
\end{listing}

\begin{listing}[ht]
\caption{{Status after unauthenticated unregistration attack on ROS Master API}}
\begin{minted}
[
frame=lines,
framesep=2mm,
baselinestretch=1.2,
fontsize=\footnotesize,
linenos
]
{bash}
root@655447dc534c:/# rosnode list
/listener
/rosout

root@655447dc534c:/# ps -e
  PID TTY          TIME CMD
    1 pts/0    00:00:00 launch_script.b
   31 pts/0    00:00:00 roscore
   42 ?        00:00:01 rosmaster
   55 ?        00:00:01 rosout
   72 pts/0    00:00:00 bash
   78 pts/1    00:00:00 bash
   90 pts/0    00:00:00 talker
  108 pts/0    00:00:01 listener
  174 pts/1    00:00:00 ps
\end{minted}    
\label{listing:attack2}
\end{listing}

\noindent As it can be seen when introspecting the ROS Master through the \emph{rosnode list} command, the \emph{/publisher} node has disappeared while the process \emph{talker} is still running (line 13 of listing \ref{listing:attack2}) and consuming resources. What happens under the hood here is that the ROS master will now notify all subscribers of the \emph{/publisher} node that it is no longer available. \\
\newline
The authors ask themselves: How can we determine whether our system has been subject of the unauthenticated unregistration attack? How can we identify at which point our robotic system was compromised? What are the affected subsystems? The following section will explore some of these questions by researching the volatile memory landscape of the presented simulated robot.


\subsubsection{Forensic study}
The study of this vulnerability will be conducted using the \emph{ros\_volatility} plugin set. In particular, the \textbf{linux\_rosnode} plugin which locates ROS nodes from memory and reports information about them. The authors will base their research on memory captures made through the LiME utility installed in their robotic system. In an untampered system, the plugin operates as follows:

\begin{listing}[ht]
\caption{{Output of running the \textbf{linux\_rosnode} plugin before the unauthenticated unregistration attack on the Master API}}
\begin{minted}
[
frame=lines,
framesep=2mm,
baselinestretch=1.2,
fontsize=\footnotesize,
linenos
]
{bash}
vagrant@vagrant-ubuntu-trusty-64:~$ vol.py 
    --plugins=/vagrant/ros_volatility 
    --profile LinuxUbuntu14045x64 
    -f robot.lime linux_rosnode
Volatility Foundation Volatility Framework 2.6
rosout
talker
listener
\end{minted}    
\label{listing:forensics1}
\end{listing}

\noindent Listing \ref{listing:forensics1} shows the result of running the \textbf{linux\_rosnode} plugin on the ROS setup, prior to the exploitation of an unauthenticated unregistration attack on the Master API. Listing \ref{listing:forensics2} displays the result of running \textbf{linux\_rosnode} after the attack (described previously in Listing \ref{listing:attack1}). Note that the \emph{talker} node has been identified as unregistered. The plugin determines this aspect by analyzing all the sockets available in memory for each one of the ROS nodes detected, and infers that those publishers with a socket in the same port both in `LISTEN` and `CLOSE\_WAIT` status were likely unregistered. The code that implements this detection is revealed at Listing \ref{listing:forensics3}.

\begin{listing}[ht]
\caption{{Output of running the \textbf{linux\_rosnode} plugin after the unauthenticated unregistration attack on the Master API}}
\begin{minted}
[
frame=lines,
framesep=2mm,
baselinestretch=1.2,
fontsize=\footnotesize,
linenos
]
{Bash}
vagrant@vagrant-ubuntu-trusty-64:~$ vol.py 
    --plugins=/vagrant/ros_volatility 
    --profile LinuxUbuntu14045x64 
    -f robot_hacked.lime linux_rosnode
Volatility Foundation Volatility Framework 2.6
rosout
talker (unregistered)
listener
\end{minted}    
\label{listing:forensics2}
\end{listing}

\begin{listing}[ht]
\caption{{Detection of unregistered nodes, snipped of code from \textbf{linux\_rosnode} plugin}}
\begin{minted}
[
frame=lines,
framesep=2mm,
baselinestretch=1.2,
fontsize=\footnotesize,
linenos
]
{Python}
# The following assumption is made for detecting 
#   unregistered nodes:
#
#     a publisher having a socket in the same 
#       port both in `LISTEN` and `CLOSE_WAIT` 
#       status was likely unregistered
#
# WARNING: this assumption was validated for a simple 
#   scenario. Further
# research needs to be executed to validate it in 
# multi-topic and multi-nodes scenarios.
listen_ports = [] # ports with LISTEN state
close_wait_ports = [] # ports with CLOSE_WAIT state
for ents in t.netstat():
    if ents[0] == socket.AF_INET:
        (_, proto, saddr, sport, daddr, dport, 
            state) = ents[1]
        if state == 'LISTEN':
            listen_ports.append(sport)
        elif state == 'CLOSE_WAIT':
            close_wait_ports.append(sport)

unregistered = False
for p in close_wait_ports:
    if p in listen_ports:
        unregistered = True
\end{minted}    
\label{listing:forensics3}
\end{listing}

\noindent By using the \textbf{linux\_rosnode} plugin, a researcher is able to determine that the \emph{talker} ROS node has been unregistered and thereby, the system, likely, compromised. A complete walk-through on the forensic study of this vulnerability is available at \url{https://github.com/vmayoral/basic_robot_cybersecurity/tree/master/robot_forensics/tutorial1}.\\
\newline
The authors now know that the system has been tampered and the talker unregistered, but what was the cause? To answer this question, one of the default volatility plugins will be used. \\
\newline
Results are displayed in Listing \ref{listing:forensics4}. Lines 23 and 24 show that it was the \texttt{roschaos} tool what exploited the unauthenticated unregistration vulnerability in the ROS Master.

\begin{listing}[ht]
\caption{{The \textbf{linux\_bash} volatility plugin recovers a history of commands, exploit discovered}}
\begin{minted}
[
frame=lines,
framesep=2mm,
baselinestretch=1.2,
fontsize=\footnotesize,
linenos
]
{bash}
vagrant@vagrant-ubuntu-trusty-64:~$ vol.py 
    --plugins=/home/vagrant/volatility-plugins/linux
    --profile LinuxUbuntu14045x64 
    -f robot_hacked.lime linux_bash
1583 bash                 2018-11-01 17:58:12 
    UTC+0000   rosnode list
1583 bash                 2018-11-01 17:58:18 
    UTC+0000   roscore &
1583 bash                 2018-11-01 17:58:28 
    UTC+0000   rosrun scenario1 talker &
1583 bash                 2018-11-01 17:58:32 
    UTC+0000   rosrun scenario1 
    listener > /tmp/listener.txt &
...
1583 bash                 2018-11-01 18:09:19 
    UTC+0000   rosnode list
1583 bash                 2018-11-01 18:09:33 
    UTC+0000   sudo insmod /lib/modules/lime.ko
    "path=/home/vagrant/robot.lime format=lime"
1583 bash                 2018-11-01 18:09:38 
    UTC+0000   sudo rmmod /lib/modules/lime.ko
1583 bash                 2018-11-01 18:09:46 
    UTC+0000   roschaos master unregister node
    --node_name /publisher
1583 bash                 2018-11-01 18:09:50 
    UTC+0000   sudo insmod /lib/modules/lime.ko 
    "path=/home/vagrant/robot_hacked.lime format=lime"
\end{minted}    
\label{listing:forensics4}
\end{listing}


\FloatBarrier

\section{Discussion and future work}
\label{conclusions}

In an attempt to provide a foundational path for the growth of robot forensics, and specifically, aimed at volatile memory, the authors presented the first steps towards reproducible robotic volatile memory forensic analyses with a focus on Linux and ROS, the \emph{de facto} standard in robotics. Common practices were introduced and a walkthrough for a simulated attack was illustrated step-by-step. Through the forensic analysis presented, the readers should be able to answer several of the proposed questions and determine the specific cause of the attack by solely looking at RAM memory.\\
\newline
Contributions have been presented in the form of a plugin to the Volatility framework. It should be noted however that due to restrictions in the framework, only (Linux) kernels below version 3.6 are supported. Future work will address this limitation and look into more elaborated forensic studies, involving more complicated attacks. Additional plugins are being developed for \emph{ros\_volatility} to detect other attacks. Eventually, the authors hope that unknown robotic attack patterns could get identified. Beyond detecting attacks, it would be of special interest the future extension of the ROS robotics framework with memory introspection tools that running side-by-side with the application, could perform real-time analyses at regular intervals to detect attacks such as the one just reported.\\
\newline
Another area of future work to approach robot forensics that could serve as a natural extension of the work presented here is the use of a dedicated tamper-proof storage device to securely store the results of the analyses done at run-time (commonly called a \"black box\"). Recently, a concept and prototype for such a device especially tailored for modern robots (and their threat profile) has been presented in \cite{Taurer2018Securedatarecording}. While currently this device needs to be specially considered within the robot application, using memory forensics to generate its input data would make it much easier applicable in existing robot software. No changes to a ROS application or the ROS core would be necessary, but still, meaningful forensic evidence on the robot system's execution could be collected. 


\appendix

\section{Robot forensics and reversing}
\label{forensicsvsreversing}

Opposed to forensics, which focuses on recovering data to establish who committed the crime\cite{ligh2014art}, reverse engineering (commonly known as reversing) is concerned with the process of extracting knowledge or design blueprints from the system\cite{eilam2011reversing}. As described by Eilam, reverse engineering is usually conducted to obtain missing knowledge, ideas, and design philosophy when such information is unavailable. In some cases, the information is owned by someone who is not willing to share. These two concepts are often mistaken. When transposed to robotics, we apply the following understanding:
\begin{itemize}
    \item \textbf{Robot forensics}: the process to obtain, preserve and document evidence (typically data) from robot-related crimes.
    \item \textbf{Robot reversing} (or robot reverse engineering): the process of extracting information about the design elements in a robotic system.
\end{itemize}

\section{Acknowledgments}

This research has been partially funded by the Basque Government, in particular, by the Business Development Basque Agency (SPRI) through the Ekintzaile 2018 program. Special thanks to BIC Araba and the Basque Cybersecurity Center (BCSC) for their support.
This work has been supported by the programme ''ICT of the Future'', managed by the Austrian Research Promotion Agency (FFG), under grant no. 861264.

\section{Availability}

All source code to reproduce the work presented here has been made publicly available. Further extensions will also be published into the same repositories to raise community interest. The authors of the paper encourage other security researchers to contribute with their own ideas an extensions. A walkthrough on the forensic study presented in this paper is also available from

\begin{center}
{\tt \url{https://github.com/vmayoral/basic_robot_cybersecurity/tree/master/robot_exploitation/tutorial11}}\\
\end{center}

\noindent additional information on the vulnerability exploited can be obtained from

\begin{center}
{\tt \url{https://github.com/aliasrobotics/RVDP/issues/87}}\\
\end{center}

\noindent finally, the extensions to Volatility to perform forensic studies can be downloaded from

\begin{center}
{\tt \url{https://github.com/aliasrobotics/ros_volatility}}\\
\end{center}








\bibliographystyle{IEEEtran}
\bibliography{bibliography}

\theendnotes

\end{document}